# Fabrication of nanopatterned metal layers on silicon by nanoindentation / nanoscratching and electrodeposition.

R. Cecchini[a]*, A. Fabrizi[a], C. Paternoster[a], W. Zhang[b], G. Roventi[b]

[a] VINF/Marche Polytechnic University, Department of Mechanics - v. Brecce Bianche, 60131 Ancona, Italy
[b] Marche Polytechnic University, Department of Physics and Material and Territory Engineering - v. Brecce Bianche, 60131 Ancona, Italy

**Abstract**

The present work illustrates a novel approach for the maskless and resistless fabrication of nanopatterned metal layers on Si substrates, based on the combination of nanomechanical surface modification techniques (such as nanoindentation and nanoscratching) and electrodeposition. Single crystal (100) *n*-doped Si substrates were first cleaned from native oxide. Nanoindentation and nanoscratching were then used to locally change the substrate microstructure and create regions with reduced electrical conductivity. The substrates were finally mounted as cathode electrodes in a three electrode electrochemical cell to potentiostatically deposit a Ni layer. Electrodeposition was prevented in regions with modified microstructure, enabling the formation of a patterned Ni layer. The fabrication of several patterns including continuous Ni lines of 200 nm width and several microns length was obtained.

**Keywords:** maskless nanopatterning; nanoindentation; electrodeposition; silicon; phase transformation.

———

* Corresponding author, tel.: +39 071 2204794, fax: +39 071 2204801, E-mail address: r.cecchini@univpm.it

## 1. Introduction

Microelectronic and micro-mechanical devices industry is in constant need of high resolution and low cost fabrication processes. One critical aspect is represented by the processes involved in the transfer of a given pattern onto a substrate. Often the substrate is constituted by a semiconductor material, like Si, onto which a patterned metal layer needs to be grown for the fabrication of micro-electro-mechanical systems (MEMS) and electronic, optoelectronic and magnetic devices. Metal nanostructures (e.g. nanometric clusters) on semiconductors can also be used as catalysts for chemical reactions and for the fabrication of sensors. Moreover, metal/semiconductor Schottky junctions have a huge number of applications (e.g. solar cells). Several processes, including electrodeposition, are studied for the fabrication of Schottky diodes [1-4].

Typically, lithographic processes such as photolithography are used for transferring the pattern from a mask to a photoresist material layer, previously deposited on the substrate. Subsequently, deposition, etching or other steps are performed in order to add and subtract material as required for transferring the pattern on the substrate [5]. Among other deposition processes, electrodeposition in combination with photolithography represents an effective and cost efficient way for fabricating, e.g., patterned magnetic metal layers [6-7]. The use of a resist for substrate patterning has the disadvantage of requiring several fabrication steps, including resist deposition, curing and removal. Several approaches for maskless and resistless nanopatterning have been attempted, including the use of nanomechanical techniques. Nanoindentation and nanoscratching are well known tools for changing as well as for measuring the properties of surfaces at nanometric scale. The combination of nanoindentation and nanoscratching with electrochemical deposition as a way for obtaining metal nanopatterning on semiconductor surfaces has been investigated by other authors [8-14]. Selective metal electrodeposition and electroless deposition within the scratches produced by an Atomic Force Microscope (AFM) or a nanoindenter in an oxide or organic insulating layer has been demonstrated [8-12]. In other works defects were created by nanoindentation on hydrogen terminated Si surface, where metal deposits were grown by electroless deposition, as a consequence of modified local surface activity [13,14]. In these types of positive lithographic processes deposition occurs on nanomechanically modified areas of the substrate.

On the other hand, nanoindentation and nanoscratching can also be used to change the microstructure, e.g. the phase, of the indented or scratched material. It is for instance known that nanoindentation and nanoscratching of single crystal Si can induce amorphisation in a limited volume of material near the surface [15-18]. The use of scratching was for instance considered for the fabrication of amorphous patterns on single crystal Si, subsequently used as an etching masking [19].

Aim of this work is to study a novel approach to the maskless and resistless fabrication of nanopatterned metal layers on Si, where nanomechanically induced local amorphisation of the substrate is used as a mask for electrodeposition. By this approach, the combination of nanomechanical surface modification and electrodeposition results in a negative type of lithography for metal nanopatterning.

Patterned Ni layers on *n*-doped single crystal (100) Si substrates were fabricated using nanoindentation and nanoscratching followed by electrodeposition and characterized by Scanning Electron Microscopy (SEM).

**2. Experimental**

Commercially available 380 μm thick *n*-type single crystal (100) Si wafers (phosphorous doped with resistivity of 1-10 Ω·cm) with one mechanically polished side were cut into 1 cm x 1 cm samples and immersed in 3.5 % HF solution for 10 min to remove native $SiO_2$ layer and leave a hydrogen terminated surface. A Hysitron® UBI nanoindenter with a Berkovich diamond tip (curvature radius ∼1 μm) was used to perform nanoindentation and nanoscratching on the Si samples top surface. Nanoindentations were performed by applying a force perpendicularly to the substrate surface (z direction) with loading and unloading sequences of 10 s and peak forces of 1000 to 10000 μN. Using its scanning probe mode, the nanoindenter was also operated for nanoscratching, by simultaneously applying a force along the z direction and scanning the tip in the lateral, x-y, directions. Nanoscratching was done at constant forces of 200, 300, 400 and 500 μN and scanning rate of 130 μm/s.

After nanomechanical surface modification step, an ohmic contact was fabricated on the Si samples back surface using InGa eutectic (99.99%). The samples were then mounted on a copper support. An insulating adhesive tape was used to mask off the support and the sample leaving exposed an area of 0.196 $cm^2$ on the Si top surface.

Si amorphisation in the scratched regions was investigated by micro-Raman spectroscopy, using a Jobin-Yvon® micro-Raman system with 632.8 nm excitation laser, and by conductive Atomic Force Microscopy (AFM), using a NT-MDT® Solver P47 Pro AFM operated with Pt coated probes.

A conventional three electrode cell was used to perform Cyclic Voltammetry (CV) and electrodeposition; a platinum spiral was the counter-electrode and an Ag/AgCl (E = +0.208 vs. SHE) was the reference electrode, to which all potentials are referred to. The reference electrode was mounted inside a Luggin capillary, whose tip was placed next to the working electrode surface. A Watts type bath was used for electrochemical tests: $NiSO_4 \cdot 6H_2O$ 1.14 M, $NiCl_2 \cdot 6H_2O$ 0.15 M, $H_3BO_3$ 0.65 M and saccharine 1 g·dm$^{-3}$. All the solutions were prepared with double-distilled water and analytical grade reagents. All electrochemical experiments were performed at room temperature (T=25 °C). Bath composition and deposition temperature were chosen in order to obtain nanocrystalline Ni deposits. In fact, preliminary tests indicated that addition of saccharine and deposition at room temperature favours the reduction of Ni grain size, as measured by Scherrer's equation, using the full-width at half maximum (FWHM) of diffraction peaks recorded from electrodeposited Ni layers with a Philips® PV 1730 X-Ray Diffractometer (XRD) with $\lambda_{Cu\ K\alpha}$=0.154 nm. Before each electrochemical experiment, solutions were deaerated by $N_2$ bubbling inside the cell. CV was performed to characterize the electrochemical system and establish Ni deposition potential on Si. CV tests were carried out using an EG&G Princeton Applied Research® potentiostat/galvanostat model 273 controlled by a personal computer; the scanning rate was 20 mV/s. Electrodeposition experiments were performed potentiostatically with deposition potential in the range of -1.000 to -3.000 V and carried out by an AMEL® model 549 potentiostat connected to an AMEL® model 568 programmable generator function for deposition times in the range 0.2 to 2 s. The current density ($j$) - time ($t$) curves measured during electrodeposition were recorded by a computer connected to the potentiostat. After electrodeposition, samples were investigated by a Philips® XL20 SEM equipped with an Edax® Energy Dispersive X-ray Spectroscopy (EDX) system.

A schematic description of the present approach for maskless and resistless metal nanopatterning on Si is illustrated in Fig. 1. After native oxide layer removal by immersion in HF solution, Si surface is modified by nanoindentation or nanoscratching, which generates amorphous regions on Si. Finally Ni electrodeposition is carried out and a Ni layer is grown on unmodified Si regions.

## 3. Results and Discussion.

First, force ($P$) - penetration depth ($h$) curves were used to characterize Si substrates behaviour during nanoindentation. A typical example of recorded $P$-$h$ curves is reported in Fig. 2. These curves are similar to those reported in the literature [15-17] and are related to the indentation parameters, to the substrate mechanical properties and to mechanically induced phase transformations in the substrate. It is known that in nanoindentation atmospheric phase Si-I (diamond structure) transforms into a metallic Si-II (b-Sn structure) phase on loading when a critical pressure (11-13 GPa) is attained under the indenter tip [20]. Other phase transformations are possible on unloading. The metallic phase can transform into a-Si or into high pressure polycrystalline phases Si-III (bc8, body-centered cubic structure) and Si-XII (r8, rhombohedral structure). These transformations are accompanied either by a discontinuity, called "pop-out", or by a so called "elbow" in the unloading part of the $P$-$h$ curve. Strong correlation has been found between the "pop-out" and the appearance of Si-III and Si-XII, while the presence of the "elbow" is linked to the appearance of a-Si [15]. It has been also proved that the above mentioned transformations on the Si indented volume can be controlled by controlling the indentation parameters: tip geometry, peak load, loading and unloading rates [21]. It was argued that Si-II→Si-XII/Si-III transformation can only occur when sufficient time for lattice reconstruction is given to the indented material. a-Si formation is therefore favoured by fast unloading. a-Si formation is also favoured when unloading is performed from low peak loads, as the volume of modified material is in such cases too low to allow a sufficient number of nucleation sites to be created for the reconstructive Si-II→Si-XII/Si-III transition. The recorded curves, such as the one reported in Fig. 2, with the presence of an "elbow" on the unloading part of the *P-h* curve, suggest that under present experimental conditions, and in particular the relatively low peak loads used, a-Si is formed during nanoindentation. Using the nanoindenter for nanoscratching, a-Si lines were also created on the Si surface. Formation of a-Si during lubricated and dry scratching has previously been observed [17-19].

Fig. 3(a) reports the results of micro-Raman investigation on a scratched and on an unmodified Si region. Subtraction of the two spectra (scratched Si and reference unmodified Si) is reported in Fig. 3(b). The comparison with Raman measurements reported in the literature [15-17] confirms the attribution of the broad bands at 160, 310, 400 and 470 cm$^{-1}$ to a-Si phase.

Electrical characterisation of the a-Si layer formed during scratching was done by conductive AFM. Fig. 4 reports the current-voltage curves measured on both scratched and unmodified Si areas. The current-voltage curve measured on unmodified crystal Si is determined by the properties of the metal/semiconductor junction formed between the Pt of the AFM tip and the crystal *n*-Si: the initial exponential increase of the current with forward voltage bias (*n*-Si substrate negative with respect to AFM tip) is compatible with the presence of a Schottky barrier at the junction. On the other hand, no forward current is observed on scratched areas up to + 3.0 V forward voltage bias; while only a low value of current is present above this value of bias.

Characteristics of the electrolyte/Si systems were investigated by CV tests, performed both for as received native oxide covered and for HF cleaned Si samples (Fig. 5). Ni electrodeposition starts at about -0.920 V on HF cleaned Si, while it starts at about - 1.300 V on native oxide covered Si. During the reverse scan Ni deposition continues also for more positive potential values; this is due to the fact that in the reverse scan Ni deposition occurs on the metallic layer deposited during the direct scan. An anodization peak, due to Ni dissolution, is observed during the reverse scan on HF cleaned Si substrates. No anodization peak is observed during the reverse scan on native oxide covered Si substrates, and SEM investigations and EDX analysis confirmed that no Ni dissolution occurs during the reverse scan on such samples. Hence, the presence of native $SiO_2$ layer on the Si substrate surface shifts Ni deposition potential towards more negative values compared to oxide free Si surface during the direct scan, as well as hindering Ni dissolution during the reverse scan. The absence of anodization during the reverse scan of the CV in presence of $SiO_2$ can be in fact explained as a result of the electric properties of the Ni/substrate junction formed during the forward scan of the CV. Current-voltage measurements on electrodeposited Ni/*n*-Si and Ni/$SiO_2$/*n*-Si samples showed a rectifying behaviour for both types of junctions, with an exponential increase of current as a function of voltage for forward bias (*n*-Si negative with respect to Ni layer, as during forward scan of the CV) and a constant value of current recorded for reverse bias (*n*-Si positive with respect to Ni layer, as during reverse scan of the CV). For Ni/*n*-Si samples, the value of measured reverse current (~ 1 mA cm$^{-2}$) is high enough to allow Ni dissolution during the reverse scan of the CV. For Ni/SiO$_2$/n-Si reverse current is extremely low (~1·10$^{-3}$ mA cm$^{-2}$) and is not sufficient to allow significant Ni dissolution during the reverse scan of the CV. Similar properties of electrodeposited metal/semiconductor junctions were reported in other works [1-4]. The effect of SiO$_2$ interlayer on the reduction of the reverse current of the junctions is also expected [5].

Ni electrodeposition was then carried out on nanomechanically modified Si substrates. Fig. 6(a) shows an SEM image of a Si sample where Ni electrodeposition was performed after nanoindentation with peak load of 10000 μN. It is seen that a continuous Ni film is formed on the unmodified Si surface, whereas no film growth takes place in the indented region, as a result of amorphisation of Si and consequent reduced electrical conductivity, as described by Fig. 4. Fig. 6(b) reports an SEM image of a sample where indentation was performed with 1000 μN. As it can be seen, also for this value of peak load amorphisation of Si was obtained. Some Ni deposition defects appear near the edges of the imprint, probably as a consequence of not complete amorphisation of Si in the indented area.

This indicates the possible use of nanomechanical modification of Si surface and electrodeposition for fabricating patterned metal layers. Incidentally, this procedure can also be useful as a complementary technique for studying nanomechanically induced phase transformation on Si. This has been the subject of several investigations, by Transmission Electron Microscopy (TEM), Raman spectroscopy and electrical conductivity measurements by conductive indentation tips and Scanning Probe Microscopy (SPM) techniques [15, 22-25]. The extension and shape near surface of the Si region with significant reduction of electrical conductivity, due to microstructural changes, can be quickly evaluated as it corresponds to the region where no electrodeposition takes place. From Fig. 6(a) and (b) it can be seen that the amorphised region follows the shape of the indenter Berkovich tip.

A typical *j-t* curve recorded during Ni electrodeposition is shown in Fig. 7. Cathodic current density rises up with *t*, reaches a maximum value and then starts falling. This is consistent with a deposition process that proceeds by nucleation followed by nuclei growth. Three dimensional nucleation and growth is often observed for metal electrodeposition on semiconductors, as a consequence of weak interaction between metal adatoms and substrate [26,27]. Film growth by coalescence of Ni nuclei is confirmed by SEM images (Fig. 6).

Using the nanoindenter for nanoscratching and by controlling the nanoindenter tip movement in the x-y directions different a-Si patterns, such as lines, squares and arrays were created on Si. Electrodeposition followed, generating a Ni patterned layer.

In Fig. 8 is reported the image of two 15 μm x 30 μm rectangular patterns separated by a 1 μm wide line, fabricated by scanning the nanoindenter tip with a constant force of 300 μN along the x direction followed by electrodeposition of a 200 nm thick Ni film. The distance between the scanned lines, 100 μm in this case, was chosen so that the amorphised lines were overlapping. This prevents Ni film to grow inside the scanned areas. Absence of Ni in the scanned area was verified by EDX measurements. It is important to notice here that the masking properties of the amorphised Si are extremely good. No Ni electrodeposition occurred inside the patterns fabricated by scratching at constant force of 300 μN for deposition potentials as negative as -3.00 V. This is a much more negative potential than the value at which electrodeposition starts on Si covered by native $SiO_2$ (Fig. 4), increasing the deposition selectivity of the present approach compared to those based on masking by native oxide. Considering that electrodeposition starts on hydrogen-terminated samples at -0.920 V, this gives a wide range of deposition potentials to be used for Ni electrodeposition. It has also to be noticed that, if needed, the amorphous Si layer produced by nanoindentation and nanoscratching can be easily removed after electrodeposition by HF solution clean [19].

The smallest force at which no Ni deposition was observed on scratches was 200 μN. It can be inferred that for such load amorphisation of Si is still obtained. In [19] diamond tips with curvature radius of 300 – 1000 nm were used for scratching a Si substrate, forming a-Si patterns. It was found that vertical forces as low as 180 μN are sufficient to create a layer of a-Si of few nanometres of thickness.

Scanning the surface along further apart lines allowed the fabrication of parallel Ni lines such as those reported in Fig. 9. Continuous and well separated lines with several microns length and submicron width were electrodeposited. The quality of the deposits obtained with the present approach seems to be improved with respect to other methods based on the scratching through an oxide layer when considering the submicrometric range [9,10]. One of the reasons of such improvement can be found in the fact that in the present approach the deposition occurs on the unmodified, and hence flat, Si surface rather than on scratched, and hence rough, regions (e.g. grooves). Moreover, in [9,10] it was reported that nucleation starts at the AFM-scratch edges and metal lines result from the coalescence of two separate lines into a single one, as a consequence Si amorphisation in the centre of the groove. This effect does not represent an issue with the present approach, as the formation of a-Si is precisely the principle exploited to obtain nanopatterning. The metal lines formed by our approach do not present this double feature characteristic.

The smallest line width obtained was below 100 nm, although these lines were discontinuous (Fig. 10). Finally, by scanning the nanoindenter in two perpendicular directions, arrays of Ni islands were electrodeposited. Fig. 11 reports the SEM image of one of such arrays with nanometric Ni deposits. The array was obtained by scanning the nanoindenter tip with a force of 300 μN along lines with 250 nm separation first in the x and then in the y direction, followed by Ni electrodeposition at -2.300 V for 0.2 s.

Further improvement on quality and scaling down of the patterning could be attained by additional investigation on the effect of scanning rate during scratching and by use of tips with different geometry. For what concerns the electrodeposition step, the effect of deposition potential and other additives besides of saccharine, on the nucleation and growth stage could be investigated in order to improve deposit quality.

4. Conclusions.

A novel maskless and resistless processing method for the fabrication of patterned metal films on Si was studied. The method combines nanomechanical surface modification techniques, for changing the substrate microstructure, and selective film growth by electrodeposition. Nanoindentation and nanoscratching were used to induce localized amorphisation on the surface of *n*-doped single crystal (100) Si substrates. The regions with modified microstructure were used to mask the substrates during Ni electrodeposition step from a Watts type bath, leading to the formation of patterned Ni layers. Several patterns with characteristic dimensions both in the micrometric and nanometric range were fabricated. As the size of the amorphous lines created during scratching depends on the applied force and on tip size, it is foreseen that by using shaper tips even smaller features could be created. Diamond nanoindentation tips with curvature radius smaller than 50 nm are already commercially available. In alternative an AFM with diamond or diamond coated tips could be used instead of a nanoindenter to attain even smaller feature sizes.

**Acknowledgments**

The authors are thankful to Dr. T. Bellezze, Prof. M. Cabibbo, Prof. R. Fratesi and Prof. S. Spigarelli for useful discussion, to Dr. A. Di Cristoforo for XRD measurements, to Dr. L. Gobbi and Dr. G. Barucca for their assistance with SEM investigations, to the Microwave and Optics Group of Electronic Engineering Dep. (Marche Polytechnic University) for conductive AFM measurements and to the Physics Dep. and C.I.G.S. Laboratory of the University of Modena and Reggio Emilia for Raman and FIB-SEM investigations.

**List of figures captions**

**Fig. 1.** Schematic description of the process steps for maskless and resistless patterning: a) *n*-doped single crystal (100) Si substrate covered with native oxide layer; b) native oxide removal by immersion in HF solution; c) Si surface nanoindentation or nanoscratching by diamond tip, which generates amorphous regions on Si; c) Ni electrodeposition on unmodified Si regions.

**Fig. 2.** Force (*P*) as function of penetration depth (*h*) plot for loading and unloading during nanoindentation of *n*-doped single crystal (100) Si substrate with a Berkovich diamond tip: peak load= 10000 µN; loading-unloading rate= 1000 µN/s. The arrow marks the "elbow" in the unloading part of the curve, indicating formation of a-Si.

**Fig. 3.** (a) Raman shift obtained from an unmodified crystal Si area (solid line) and on an area after nanoscratching with a constant force of 300 µN (dotted line); (b) Raman shift obtained by subtraction of Raman spectra obtained from unmodified crystal Si area to the Raman spectra obtained from the scratched area. The arrows mark the bands corresponding to a-Si.

**Fig. 4.** Current-voltage measurement by conductive AFM recorded on an unmodified crystal Si area (-■-) and on an area scratched with a constant force of 300 µN (-□-). Positive voltage bias corresponds to AFM tip at a positive potential with respect to Si substrate.

**Fig. 5.** Cyclic voltammogram obtained for as received (dashed line) and HF cleaned (solid line) *n*-doped (100) Si samples in the Watts type bath. Scanning rate= 20 mVs$^{-1}$.

**Fig. 6.** SEM image of a Si sample after: (a) nanoindentation with peak load= 10000 µN and loading-unloading rate= 1000 µN/s followed by Ni

electrodeposition at -1.200 V for 4.0 s; (b) nanoindentation with peak load of 1000 µN and loading-unloading rate= 1000 µN/s followed by Ni electrodeposition at -2.000 V for 0.6 s (b).

**Fig. 7.** Current density transient (*j-t*) recorded during Ni electrodeposition from the Watts type bath at deposition potential of -1.600 V.

**Fig. 8.** SEM image of two rectangular patterns in a 200 nm thick Ni film separated by 1 µm wide Ni line, fabricated by scanning the nanoindenter tip with a constant force of 300 µN along lines separated by 100 nm and Ni electrodeposition performed with potential of -1.300 V for 3.0 s.

**Fig. 9.** SEM image at low (a) and high (b) magnification of parallel 200 nm wide and 100 nm thick Ni lines fabricated by scanning the nanoindenter with a constant force of 300 µN along lines separated by 250 nm and Ni electrodeposition performed with potential of -2.000 V for 0.6 s.

**Fig. 10.** SEM image of parallel Ni lines with width lower than 100 nm. Nanoscratching was done by scanning the nanoindenter with constant force of 500 µN along parallel lines separated by 250 nm. Ni electrodeposition was done at a potential of -2.300 V for 0.2 s.

**Fig. 11.** SEM image of an array of Ni deposits with lattice period of 250 nm, obtained by scanning the nanoindenter with a constant force of 300 µN along two perpendicular directions and Ni electrodeposition performed with a deposition potential of -2.300 V for 0.2 s.